\documentstyle[aps,epsf,psfig,prb,twocolumn]{revtex}
\def\gz{\ifmmode{Z\hskip -4.8pt Z}
\else{\hbox{$Z\hskip -4.8pt Z$}}\fi} 

\newcommand{\be}{\begin{equation}}
\newcommand{\ee}{\end{equation}}
\newcommand{\bea}{\begin{eqnarray}}
\newcommand{\eea}{\end{eqnarray}}

\begin{document}
\tighten
\draft
\title{Phase diagrams of spin ladders with ferromagnetic legs}

\author{ T. Vekua $^{a}$, G.I. Japaridze $^{a,b}$, and H.-J. Mikeska$^{a}$ }
\address{$^{a}$ Institut f\"{u}r Theoretische Physik,
Universit\"{a}t Hannover, 30167 Hannover, Germany\\ 
$^{b}$ Andronikashvili Institute of Physics, Tamarashvili 6, 380077, 
Tbilisi, Georgia}
\address{~
\parbox{14cm}{\rm 
\medskip
The low-temperature properties of the spin $S=1/2$ ladder with
anisotropic ferromagnetic legs are studied using the continuum limit
bosonization approach. The weak-coupling ground state phase diagram of
the model is obtained for a wide range of coupling constants and
several unconventional gapless ''spin-liquid'' phases are shown to
exist for ferromagnetic coupling. The behavior
of the ladder system in the vicinity of the ferromagnetic instability
point is discussed in detail.
\vspace{3mm}
 }}	

\maketitle

\section{Introduction}

A theoretical understanding of the properties of quantum spin ladder
systems has attracted a lot of current interest for a number of
reasons: on the one hand there is an increasing number of new magnetic
materials with a ladder-like structure characterized by rich ground
state phase diagrams \cite{RiceDagotto}. On the other hand,
spin-ladder models pose interesting theoretical problems. For example,
since a spin-$S$ chain can be described as a $2S$-leg ladder with spin
$S=1/2$, provided the interchain coupling is appropriately chosen
\cite{Schulz1,Affleck1,Nersesyan96} the even- or odd-leg ladder
systems are an excellent demonstration for Haldane's conjecture
\cite{Haldane} as generalized to $S=1/2$ ladders: the {\em
antiferromagnetic} spin ladder with an even number of legs corresponds
to a spin chain with integer spin and is predicted to have a gap,
while a ladder with odd number of legs has a gapless excitation
spectrum. The two-leg antiferromagnetic ladder is presumably the
simplest spin system which allows to follow the continuous evolution
between spin $S=1/2$ and $S=1$ antiferromagnetic chains nearly exactly
\cite{Nersesyan96,KolezhukMikeska97}.

The two-leg ladder model has been the subject of considerable
theoretical interest
\cite{DagottoMoreo,Hida,StrongMillis1,DagRieraScalap,Watanabe1,Watanabe2,Barnes1,Gopalan1,Gopalan2,Noack,WhiteNoakScalap,Poilblanc,Barnes2,StrongMillis2,TotsukaSuzuki,White96,KolezhukMikeska98,KolezhukMikeska,RojiMiyashita,Nersesyan97,Solyom97,Nersesyan98,Honeker,GT,Solyom99,Nersesyan00,Solyom00,Solyom01,MVM},
most of this work, however, concentrated on isotropic or weakly
anisotropic {\em antiferromagnetic} chains coupled by an interchain
exchange of arbitrary sign.  Ladder systems with {\em ferromagnetic
legs} are much less studied although, as will be shown here they
exhibit new interesting aspects.  The ferromagnetic ladder model in
the vicinity of the ferromagnetic instability point was recently
studied by Kolezhuk and Mikeska within the framework of quasiclassical
analysis based on a nonlinear $\sigma$-model approach
\cite{KolezhukMikeska}. In this work we extend these studies
addressing the problem of ref. \onlinecite{KolezhukMikeska} in the
extreme quantum limit of spins 1/2 and investigate the weak-coupling
ground state phase diagram of a $S=1/2$ ladder system with
ferromagnetic legs and anisotropic interleg exchange using the
continuum limit bosonization approach.

The Hamiltonian of the model under consideration is given by 
\begin{equation}\label{Hamiltonian}
\hat{H} = H_{leg}^{1} + H_{leg}^{2} +  H_{\perp}\, ,
\ee
where the Hamiltonian for leg $\alpha$ is
\bea\label{FerroLadderHamiltonian}
H_{leg}^{\alpha} &=& -J \sum_{j=1}^N 
\big( S^{x}_{\alpha,j}S^{x}_{\alpha,j+1} +
S^{y}_{\alpha,j}S^{y}_{\alpha,j+1} \nonumber\\ 
&+& \Delta \,S^{z}_{\alpha,j}S^{z}_{\alpha,j+1} \big) \, ,
\eea
and the interleg coupling is given by
\bea\label{InterLegCoup}
H_{\bot} &=& J^{xy}_{\bot} \sum_{j=1}^N \left(S^{x}_{1,j}S^{x}_{2,j} 
+ S^{y}_{1,j}S^{y}_{2,j}\right) \nonumber\\
&+& J^{z}_{\bot}\sum_{j=1}^N S^{z}_{1,j}S^{z}_{2,j}\, . 
\eea

Here $S^{x,y,z}_{\alpha,j}$ are spin $S=1/2$ operators at the $j$-th
rung, the index $\alpha=1,2$ denotes the ladder legs. The intraleg
coupling constant is ferromagnetic, $J>0$, and therefore the limiting
case of {\em isotropic ferromagnetic} legs corresponds to
${\Delta}=1$, while the case of {\em isotropic antiferromagnetic} legs
is obtained for ${\Delta}=-1$. Since we study the ground state phase
diagram of the ladder system with ferromagnetic legs we will restrict
ourselves to consider ${\Delta} > 0$, including $|{\Delta}| \ll 1$ in
the limit of legs with strong in-plane anisotropy.

The outline of the paper is as follows: In section II we derive the
bosonized formulation of the model in the continuum limit.  In section
III we discuss the weak coupling phase diagrams of the model for three
different cases of anisotropic interleg coupling. Finally, we conclude
and summarize our results in section IV.  In the appendix we present
the spin-wave approach to study the transition line related to the
ferromagnetic instability in the system.

\section{Bosonization} 

In this section we derive the low-energy effective field theory of the
lattice model Eq.\ (\ref{Hamiltonian}). Since the weak-coupling
bosonization approach to the ladder models is based on the
perturbative treatment of the interchain couplings
\cite{Watanabe2,Nersesyan96} we start from the bosonization of
separate spin $S=1/2$ ferromagnetic $XXZ$ chains.

\subsection{Separate chains}              

The anisotropic spin $S=1/2$ Heisenberg chain with $|\Delta| < 1$ is
known to be critical. The long wavelength excitations are described by
the standard gaussian theory \cite{LP} with hamiltonian
\begin{equation}\label{SpinChainBosHam}
{\cal H}_{leg} =  {u \over 2}\int dx \, [(\partial_x \phi)^{2} +  (\partial_x \theta)^{2}].
\end{equation}
$\phi(x)$ and $\theta(x)$ are dual bosonic fields, $\partial_t \phi =
u \partial_x \theta $, and satisfy the following commutational
relation
\begin{eqnarray}
\label{regcom}
&& [\phi(x),\theta(y)]  = i\Theta (y-x)\,,  \nonumber\\ 
&& [\phi(x),\theta(x)]  =i/2\, .
\end{eqnarray}
$u$ stands for the velocity of spin excitation and is fixed 
from the Bethe ansatz solution as \cite{LP}
\bea
u &=& J\frac{K}{2K-1}\sin{\left(\pi/2K\right)}\,,
\label{u}
\eea
where $K$ is Luttinger liquid parameter known from comparison with the
exact solution of the $XXZ$ chain:
\bea
K &=& \frac{\pi}{2\arccos\Delta}\, .
\label{K}
\eea

Thus the parameter $K$ increases monotonically along the $XXZ$
critical line $ -1 < \Delta < 1$ from its minimal value $K=1/2$ at
$\Delta =-1$ (isotropic antiferromagnetic chain), is equal to unity at
$\Delta =0$ (the $XY$ chain) and $K \to \infty$ at $\Delta \to 1$. At
$\Delta = 1$ the spin excitation velocity vanishes, $u=0$. This
corresponds to the {\em ferromagnetic instability} point of a single
chain.

To obtain the bosonized version of the ladder Hamiltonian we need the
explicit bosonized expressions of the spin operators. The bosonization
procedure for the spin $S=1/2$ Heisenberg chain is reviewed in many
places \cite{Affleck,Shankar,GNT}. However, since we consider the
ladder model with {\em ferromagnetic legs}, our bosonization
conventions require some comments. The unitary transformation
\be\label{UnitTrans1}
S_{\alpha,j}^{x,y}\rightarrow (-1)^{j}S_{\alpha,j}^{x,y}\, , \quad 
S_{\alpha,j}^{z} \rightarrow S_{\alpha,j}^{z}\,
\ee
changes the sign of the intrachain transverse exchange and maps the
Hamiltonian ~(\ref{Hamiltonian}) to the Hamiltonian with {\em
antiferromagnetic} legs. To maintain the ferromagnetic character of
the in-plane correlations in the bosonization, it is convenient to
implement the multiplicative factor $(-1)^{j}$, introduced by the unitary
transformation ~(\ref{UnitTrans1}), directly in the bosonized
expressions for the transverse components of the spin operators. Using
the standard bosonization formulas \cite{Affleck,Shankar,GNT} we
obtain:
\begin{eqnarray}
S_{j,\alpha}^{x} &\simeq & \frac{{\it c}}{\sqrt{2\pi}} 
:\cos\sqrt{\frac{\pi}{K}} \theta_{\alpha}: \,\nonumber\\ 
+ &(-1)^j & \, \frac{{\it ib }}{\sqrt{2\pi}}
:\sin\sqrt{4\pi K}\phi_{\alpha} \sin \sqrt{\frac{\pi}{K}} \theta_{\alpha} :\, ,
\label{bosforSx}\\
S_{j,\alpha}^{y}  &\simeq &  \frac{{\it  c}}{\sqrt{2\pi}}
:\sin\sqrt{\frac{\pi}{K}} \theta_{\alpha}:\, \nonumber\\ 
-&(-1)^j&\, \frac{{\it ib}}{\sqrt{2\pi}} 
:\sin\sqrt{4\pi K}\phi_{\alpha}\cos\sqrt{\frac{\pi}{K}} \theta_{\alpha}:\, ,
\label{bosforSy}\\
S_{j,\alpha}^{z} &=&  \sqrt{\frac{K}{\pi}} \partial_x \phi_{\alpha}\, + \nonumber\\ 
&(-1)^j&  \,
 \frac{{\it a}}{\pi} :\sin\sqrt{4\pi K}\phi_{\alpha} (x): \, .
\label{bosforSz}
\end{eqnarray}

Note that the first and the second terms in Eqs.\
(\ref{bosforSx}),(\ref{bosforSy}) are hermitian because of
(\ref{regcom}).  Furthermore $:...:$ denotes the normal ordering with
respect to free bose system (\ref{SpinChainBosHam}), $\alpha$ is the
leg index. The non-universal real constants {\it a}, {\it b} and {\it
c} depend smoothly on the parameter $\Delta$, are of the order of
unity at $\Delta=0$ \cite{Lukyanov,Hikihara} and are expected to be
nonzero at arbitrary $\Delta < 1$.

\subsection{Coupled Spin-1/2 Chains} 

For coupled $S=1/2$ chains we start with two critical $S=1/2$
Heisenberg chains and treat the interleg coupling as a perturbation
assuming $ |J^{z}_{\bot}|,|J^{xy}_{\bot}| \ll J$.  Therefore we start
with two free bose field Hamiltonians (\ref{SpinChainBosHam}) and
simply attach a leg index $\alpha=1,2$ to the fields.

Then we introduce the symmetric and antisymmetric combinations of the
bosonic fields:
\begin{equation}\label{pmFields}
\phi_{\pm}=\sqrt{\frac{1}{2{\Lambda_{\pm}} }}\left(\phi_1\pm \phi_2\right)\, , 
\theta_{\pm}= \sqrt{\frac{{\Lambda_{\pm}}}{2}}\left(\theta_1 \pm \theta_2\right), 
\end{equation}
where
$$
\Lambda_{\pm}=\left( 1 \mp \frac{1}{2\pi}\frac{J^{z}_{\bot}}{J_{eff}} \right) 
$$
and 
$$
J_{eff}  =  J \frac{1}{2K-1}\sin{\frac{\pi}{2K}}\, .
$$

Using Eqs. (\ref{bosforSx})-(\ref{bosforSz}), we finally obtain the
following bosonic Hamiltonian density:
\begin{eqnarray}
{\cal H} & = &{\cal H}^{+}+{\cal H}^{-} + {\cal H}^{\pm}_{int}\, ,
\label{FullHam}\\
{\cal H}^{+} & =& {u_{+} \over 2} [(\partial_x \theta_{+})^{2} + (\partial_x \phi_{+})^2] \nonumber\\
&-& \frac{{\cal J}_{\bot}^{z}}{2\pi}\cos{\sqrt {8 \pi K_{+}}\phi_{+}(x)}\, ,
\label{SG+}\\
{\cal H}^{-} & = &  {u_{-} \over 2}[(\partial_x \theta _{-})^{2} 
+ (\partial_x \phi_{-}(x))^2] \nonumber\\
&+& \frac{{\cal J}_{\bot}^{z}}{2\pi} \cos{\sqrt{8 \pi K_{-}}\phi_{-}(x)}\nonumber\\ 
&+&  \frac{{\cal J}_{\bot}^{xy}}{2\pi}\cos{\sqrt{\frac{2 \pi}{K_{-}}}\theta_{-}(x)}\, ,
\label{SG-}\\
{\cal H}^{\pm}_{int}  &=& \frac{{\cal J}_{+-}}{2\pi}\, 
\cos{\sqrt{\frac{2 \pi}{K_{-}}}\theta_{-}(x)}\cos{\sqrt {8 \pi K_{+}}\phi_{+}(x)}\, .
\label{SGint}\
\end{eqnarray}
Here 
\bea
u_{\pm} &=& \frac{u}{\Lambda_{\pm}} \simeq  u \left( 1 \pm \frac{1}{2\pi}\frac{J^{z}_{\bot} }{J_{eff}} \right)
\label{upm}\\
K_{\pm} & = & K \cdot \Lambda_{\pm} \simeq K 
\left( 1 \mp \frac{1}{2\pi}\frac{J^{z}_{\bot}}{J_{eff}} \right) \, , 
\label{Kpm}
\eea
and we have introduced the following coupling constants:
\begin{eqnarray}
{\cal J}_{\bot}^{z}  = J_{\bot}^{z}/\pi\, ,  \label{constant}
\end{eqnarray} for $\Delta=0$ and otherwise

\begin{eqnarray}
{\cal J}_{\bot}^{z}  &\sim& J_{\bot}^{z}\, , 
\label{Jzz}\\ 
{\cal J}_{\bot}^{xy},{\cal J}_{+-} &\sim& J_{\bot}^{xy}\, , 
\label{Jxy1}
\end{eqnarray}
with some positive constants of proportionality which cannot be fixed
by symmetry arguments (contrary to the constant appearing in Eq. 
(\ref{constant})).

In deriving (\ref{FullHam}), a term $\sim \cos{\sqrt{\frac{2
\pi}{K_{-}}}\theta_{-}}\cos{\sqrt {8 \pi K_{-}}\phi_{-}}$ which is
strongly irrelevant at $\Delta > 0 $ (ferromagnetic legs) was omitted.
Thus our approach is tailored to cover ferromagnetic intraleg coupling 
and can be applied to antiferromagnetic intraleg coupling only for
$\vert \Delta \vert \ll 1$.

\subsection{The effective continuum-limit model} 

At $J^{z}_{\bot}=J^{xy}_{\bot}=0$ the Hamiltonian (\ref{FullHam})
describes two independent Gaussian fields, i.e.~two gapless fields,
each describing a critical spin S=1/2
Heisenberg chain. Let us first address the question whether the
interleg exchange leads to the dynamical generation of a gap in the
excitation spectrum. In the case of anisotropic interleg exchange it
is rather instructive to study the effect of the longitudinal
($J_{\bot}^{z}$) and transverse ($J_{\bot}^{xy}$) part of the interleg
coupling separately.

At $J_{\bot}^{xy}=0$ the effective theory of the original ladder model
is given by two decoupled quantum sine-Gordon models
\be\label{EffectiveHamiltonianJzz}
{\cal H}_{eff}   =  {\cal H}^{+} + {\cal H}^{-}
\ee
where
\begin{eqnarray}\label{SG+-}
{\cal H}^{\pm} & = & u_{\pm}\int dx \,\Big[ {1 \over 2}[(\partial_{x} \theta_{\pm}(x))^{2}
+ (\partial_x \phi_{\pm}(x))^2] \nonumber\\
&+& {M_{\pm} \over 2\pi}  \cos{\sqrt {8 \pi K_{\pm}}\phi_{\pm}(x)} \Big] \, .
\end{eqnarray}
The two SG models respectively describe the symmetric ($\phi_{+}$) and
antisymmetric ($\phi_{-}$) degrees of freedom.

The bare values of the dimensionless coupling constants $M_{\pm}$ and
$K_{\pm}$ are known only in the weak-coupling limit $|J_{\bot}^{z}|/J,
|\Delta | \ll 1$ where they have the values:
\begin{eqnarray}
M_{\pm}  &=&  \mp {J_{\bot}^{z} \over \pi J}\, , 
\label{M}\\
K_{\pm}  &=& 1 + \frac{2\Delta}{\pi} \mp \frac{J_{\bot}^{z}}{2 \pi J}\, .
\label{K1}
\end{eqnarray}

The scaling dimensions of the {\em cosine} terms in (\ref{SG+-}) are
$d^{z}_{\pm} = 2K_{\pm} \simeq 2$. Therefore, in the weak-coupling
limit, both SG models have marginal dimension and details of their
behavior should be determined within the framework of the
renormalization-group analysis.  However, a rather straightforward
estimate indicates that at $\Delta \gg 4J_{\bot}^{z}/J$ we have
$K_{\pm} >1$ and the {\em cosine} terms are irrelevant. Therefore in
this case one concludes that the effective model reduces to the theory
of two independent Gaussian fields $\phi_{\pm}$. The effect of the
interleg coupling is extremely weak and is completely absorbed in the
renormalized values of the spin-liquid parameters $K_{\pm}$
characterizing respectively gapless symmetric and antisymmetric
modes. However, it is important to note, that in the vicinity of the
single chain ferromagnetic instability point, at $\Delta \to 1$, the
effective bandwidth collapses, $J_{eff} \simeq 2(1 -\Delta) \to
0$. Therefore in this limit the effect of the interleg coupling
becomes very strong. This implies subtle effects to be discussed
later.

The transverse interleg exchange ($J_{\bot}^{xy}$) leads to the
appearance of the {\em strongly relevant} operator ${\cal
J}_{\bot}^{xy}\cos{\sqrt{2 \pi K^{-1}_{-}}\theta_{-}}$ with the
scaling dimension $d^{xy}_{-} = (2 K_{-})^{-1} \leq 1/2$ in the
theory. Therefore, the {\em antisymmetric sector is gapped at
arbitrary} $J_{\bot}^{xy} \neq 0$. Fluctuations of the field
$\theta_{-}(x)$ are completely suppressed in this sector and
$\theta_{-}(x)$ is condensed in one of its vacua. The vacuum
expectation value of the {\em cosine} term is
\begin{equation}
\langle\cos{\sqrt{2 \pi K^{-1}_{-}}\theta_{-}}\rangle = \gamma
\end{equation}
with $\gamma \sim \left(|{\cal J}_{\bot}^{xy}| /J_{eff}
\right)^{\frac{d^{xy}_{-}}{2-d^{xy}_{-}}} \ll 1$ in weak coupling
and is of the order of unity at $|{\cal J}_{\bot}^{xy}| \geq J$.

Therefore, the condensation of the field $\theta_{-}$ strongly
influences the coupling between the symmetric and antisymmetric modes
induces by ${\cal H}^{\pm}_{int}$. Taking into account that the
fluctuations of the field $\theta_{-}$ are stopped, one easily finds
that at $J_{\bot}^{xy} \neq 0$ infrared behavior of the symmetric
field is governed by the following ''effective'' sine-Gordon theory
\begin{eqnarray}\label{SGeffective}
{\cal H}^{+}_{eff} & = & u_{+}\int dx \,\Big[ {1 \over 2}[(\partial_{x} \theta_{+})^{2}
+ (\partial_x \phi_{+}(x))^2] \nonumber\\
&+& {M^{+}_{eff} \over 2\pi}  \cos{\sqrt {8 \pi K_{+}}\phi_{+}(x)} \Big] \, ,
\end{eqnarray}
where 
\be
M^{+}_{eff} =  -\frac{1}{\pi u_{+}}\left({\cal J}^{z}_{\bot} + \gamma \cdot 
|{\cal J}_{+-}|\right)\, .
\ee
This mapping of the initial spin $S=1/2$ ladder model onto the quantum
theory of Bose fields described in terms of an ''effective''
sine-Gordon (SG) Hamiltonian (\ref{SG+-}) or (\ref{SGeffective}) will
allow to extract the ground state properties of the $S=1/2$ ladder
using the far-infrared properties of the quantum SG theory.

\subsection{The RG analysis} 

The infrared behavior of the SG Hamiltonian is described by the
corresponding pair of renormalization group (RG) equations for the
effective coupling constants ${\cal K}(l)$ and ${\cal M}(l)$
\begin{eqnarray}\label{RGeq}
\frac{d{\cal M}(l)}{dl} &=& -2\left({\cal K}(l)-1\right) {\cal M}(l)\nonumber\\
\frac{d{\cal K}(l)}{dl} &=& - \frac{1}{2}{\cal M}^{2}(l)
\end{eqnarray}
where $l=\ln(a_{0})$ and the bare values of the coupling constants are
${\cal K}(l=0) \equiv K$ and ${\cal M}(l=0)\equiv M $. The pair of RG
equations (\ref{RGeq}) describes the Kosterlitz-Thouless transition
\cite{KT}. The flow lines lie on the hyperbola
\be\label{hyperbola}
4\left({\cal K} - 1\right)^{2}-{\cal M}^{2} = \mu^{2} = 4(K-1)^{2}-M^{2} 
\ee
and exhibit two different regimes depending on the relation between
the bare coupling constants - \cite{Wiegmann} (see
Fig.\ref{fig:RGFlow}):
\begin{figure}
\vspace{10mm}
\centerline{\psfig{file=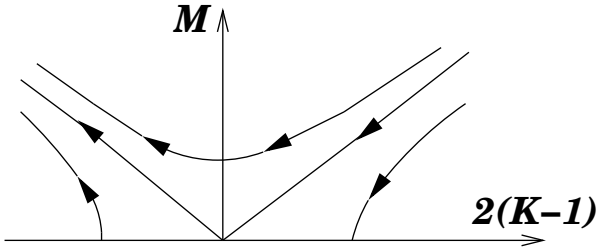,width=80mm,silent=}}
\vspace{5mm}
\caption{Renormalization-group flow diagram; the arrows denote the 
direction of flow with increasing length scale. }
\label{fig:RGFlow}
\end{figure}

{\em Weak coupling regime.} \ \ For $2(K-1)\geq \left|M\right|$ we are
in the weak coupling regime: the effective mass ${\cal M} \to 0$. The
low energy (large distance) behavior of the corresponding gapless mode
is described by a free scalar field.

The vacuum averages of exponentials of the corresponding fields show
a power-law decay at large distances
\be
\langle \, e^{i K \phi(0)} \, e^{ -i K \phi(r)}\rangle  \sim  
\langle \, e^{ i K \theta(0)} \, e^{ -i K \theta(r)}\rangle
\sim \left| r \right|^{-\frac{K^{*2}}{2\pi}}\, .
\label{FREECORRELATIONS}
\ee
where $K^{\ast}$ is the fixed-point value of the parameter $K$ determined from the 
Eq.~(\ref{hyperbola}).

{\em Strong coupling regime.} \ \ For $2(K-1) < \left|M\right|$ the
system scales to strong coupling: depending on the sign of the bare
mass $M$, the renormalized mass ${\cal M}$ is driven to $\pm\infty$,
signaling a crossover to one of two strong coupling regimes with a
dynamical generation of a commensurability gap in the excitation
spectrum. The flow of $\mid {\cal M} \mid$ to large values indicates
that the ${\cal M}\mbox{cos}\sqrt{8\pi K}\phi $ term in the
sine-Gordon model dominates the long-distance properties of the
system. Depending on the sign of the mass term, the field $\phi$ gets
ordered with the expectation values
\begin{equation}  \label{ORDERFIELDS}
\langle\phi\rangle =\left\{ 
\begin{array}{ll}
\sqrt{\pi/8K} \hskip0.5cm \textrm{ at $M  > 0$}  \\ 
0 \hskip1.5cm \textrm{ at $M  < 0$}
\end{array}
\right. \, .
\end{equation}
Using this analysis for the excitation spectrum of the SG model and
the behavior of the corresponding fields,
Eqs. (\ref{FREECORRELATIONS}, \ref{ORDERFIELDS}), we will now discuss
the {\em weak-coupling} phase diagram of the spin $S=1/2$ {\em
ferromagnetic ladder} model Eq. (\ref{Hamiltonian}).

\section{Phase diagrams} 

In this section we discuss separately the ground state phase diagram
of the {\em ferromagnetic ladder} coupled only by the longitudinal
part of the interleg spin exchange (subsection A), coupled only by the
transverse part of the interleg spin exchange (subsection B) and by an
isotropic interleg coupling (subsection C). At this point we note that
from the structure of the interaction Hamiltonian
Eq. (\ref{InterLegCoup}) follows that the phase diagrams for case (A)
and case (B) will be symmetric with respect to the lines $J_{\perp} =
0$ since a change of sign in $J_{\perp}$ leads to a unitary equivalent
Hamiltonian. This is in contrast to case (C) where this unitary
equivalence does not exist.

\subsection{Chains coupled by the longitudinal part of the interleg exchange}

In this subsection we consider the weak-coupling phase diagram of the
spin S=1/2 ferromagnetic ladder model ~(\ref{Hamiltonian}) coupled by
a weak longitudinal intrachain exchange ($J^{xy}_{\bot}=0,
J^{z}_{\bot}\neq 0$). The bosonized version of the model ${\cal
H}_{eff}={\cal H}^{+}+{\cal H}^{-}$ where ${\cal H}^{\pm}$ are given
by Eq. (\ref{SG+-}) and the bare values of the corresponding
dimensionless coupling constants are given by (\ref{M})-(\ref{K1}). By
inspection of the initial values of the coupling constants one easily
finds that:
\begin{itemize}
\item{At $\Delta <0$ both the 
symmetric and the antisymmetric sectors are gapped (except for 
$J_{\bot }^{z} = 0)$}.
\item{At $\Delta >0$  the {\em symmetric 
sector is gapped} for $J_{\bot }^{z}/J > 2\Delta >0$ while the 
{\em antisymmetric sector is gapped} for $J_{\bot }^{z}/J <- 2\Delta <0$.} 
\end{itemize}

This determines the following three distinct sectors of the phase
diagram as traced already in the RG analysis (see also Fig. 2 below):
\begin{itemize}
\item{Sector A: $\Delta < 0$ corresponds to the phase with gapped
excitation spectrum;}
\item{Sector B: $\Delta > 0$ and $|J_{\bot }^{z}| > 2 J \Delta $
corresponds to the phase characterized by the one gapless and one
gapped mode in the excitation spectrum. In particular at $J_{\bot
}^{z} > 0$ the symmetric mode is gapped whereas the antisymmetric mode
is gapless and vice-versa at $J_{\bot }^{z} < 0$;}
\item{Sector C: $\Delta > 0$ and $|J_{\bot }^{z}| < 2 J \Delta $
corresponds to the phase where both modes are gapless.}
\end{itemize}

As we show below, the same phases are present in the strong coupling
regime.  The only phase which is missed in the weak-coupling RG
analysis is the ferromagnetic phase; it appears only in the strong
coupling regime at $\Delta \simeq 1$ or at $\Delta \ll 1$ but
$|J_{\bot }^{z}| \sim 1/\Delta \gg 1$.

To clarify the symmetry properties of the ground states of the system
in the different sectors we study the large-distance behavior of the
longitudinal
\be\label{Spinzz}
K^{zz}_{\alpha\beta}(r) :=
\langle S_{\alpha}^{z}(0)S_{\beta}^{z}(r) \rangle \, ,
\ee
and the transverse 
\be\label{Spinxy}
K^{xy}_{\alpha\beta}(r) :=
\langle S_{\alpha}^{+}(0)S_{\beta}^{-}(r) \rangle \, ,
\ee
spin-spin correlation functions for intraleg ($\alpha=\beta $) and
interleg ($\alpha \neq \beta $) spin pairs.

Using the results for the excitation spectrum and the behavior of the
corresponding fields in the gapless and gapped phases, Eqs.\
(\ref{FREECORRELATIONS})-(\ref{ORDERFIELDS}), and the expressions for
the corresponding correlation functions from bosonization, we now
discuss the characteristics of the various phases in the different
sectors of the {\em weak-coupling} ground state phase diagram.

In the sector A ($J^{z}_{\bot}<0$) the vacuum expectation values of the
fields are: $\langle \phi_{+}\rangle =
\sqrt{\displaystyle{\pi/8K_{+}}} $ and $\langle \phi_{-}\rangle = 0$.
Ordering of the $\phi_{\pm}$ suppresses transverse spin correlations,
while the longitudinal correlations are given by
$$
K^{zz}_{\alpha\beta}(r) \sim (-1)^{r}\,\cdot\mbox{\it constant}\, .
$$
Therefore at $\Delta < 0$ and $J_{\bot }^{z}<0$, the long-range
ordered (LRO) antiferromagnetic phase with inphase spin ordering on
the rungs is realized in the ground state of the system.

In the Sector A1 ($J^{z}_{\bot}<0$) the vacuum expectation values of
the fields are given by $\langle \phi_{+}\rangle = 0 $ and $\langle
\phi_{-}\rangle =\sqrt{\displaystyle{\pi/8K_{-}}} $.  This immediately
implies that in this sector
$$
K^{zz}_{\alpha\beta}(r) \sim (-1)^{\alpha +\beta} \cdot (-1)^{r}\,\cdot \mbox{\it constant}\, .
$$
Therefore, at $\Delta < 0$ and $J_{\bot}^z > 0$ the LRO
antiferromagnetic phase with antiphase intrarung spin ordering is
realized in the ground state of the system.

In the sector B (B1) the antisymmetric (symmetric) field is gapped
with the vacuum expectation value $\langle \phi_{-}\rangle = 0$
($\langle \phi_{+}\rangle = 0$). However, as can be seen from weak
coupling RG analysis as well as from the strong coupling effective
spin $1/2$ model (see below), at $J^{z}_{\bot} \neq 0$ the line
$\Delta=0$ is the phase transition line along which the gapped at
$\Delta <0$ symmetric (antisymmetric) mode becomes gapless. Therefore
in the sector B (B1) the gapless degrees of freedom corresponding to
the symmetric (antisymmetric) mode are described by the free Bose
field system with the fixed-point value of the parameters
$K^{\ast}_{\pm}$. Using Eq. (\ref{hyperbola}) and the bare values of
coupling constants (\ref{M})-(\ref{K1}) it is straightforward to show
that at $|\Delta|,|J^{z}_{\bot}/J| \ll 1$
$$
K^{\ast}_{\pm} \simeq 1 + \frac{1}{\pi}\sqrt{2\Delta(2\Delta \mp J^{z}_{\bot}/J)}\, .
$$
Note that at $\Delta=0$ the fixed-point values of the spin-liquid
parameters are $K^{\ast}_{\pm}=1$ while at $J^{z}_{\bot}=0$ (see
Eq. (\ref{Kpm})) $K^{\ast}_{\pm}=K$. Therefore we conclude that along
the line $\Delta=0$ the gapless sector in the system is identical to
a single isotropic spin $S=1/2$ Heisenberg chain, while along the line
$J^{z}_{\bot}=0$ we reach the limit of two decoupled spin $S=1/2$
Heisenberg chains.

The very existence of a gapped excitation mode accompanied with
ordering of the field $\phi_{-}$ (or $\phi_{+}$) implies suppression
of the transverse correlations. On the other hand the presence of
the gapless excitation mode leads to the power law decay of the
longitudinal spin correlations. Therefore we obtain, that in the
sector B
$$
K^{zz}_{\alpha\beta}(r) \simeq \frac{ K^{\ast}_{+}}{2\pi r^{2}}  + 
\frac{(-1)^{r}}{r^{K^{\ast}_{+}}}\, ,
$$
while in the sector B1
$$
K^{zz}_{\alpha\beta}(r) \simeq (-1)^{\alpha+\beta}\, 
\left[ \frac{ K^{\ast}_{-}}{2\pi r^{2}}  + \frac{(-1)^{r}}{r^{K^{\ast}_{-}}}\right]\, .
$$
We denote this phase as the {\em spin liquid I} phase. It is interesting to
note that in sector B the following operator shows quasi long range
behavior:

\begin{eqnarray}
\langle (S_1^+(r) &+& S_2^+(r))^2 \, (S_1^-(0) + S_2^-(0))^2 
         \rangle \nonumber \\
&\simeq& 4 \langle S_1^+(r) S_2^+(r)) \, S_1^-(0) S_2^-(0) \rangle \nonumber \\
&\simeq& r^{-1/K_+^*} \,\, (-1)^r \, r^{-1/K_+^* - K_+^*}. 
\end{eqnarray}
$(S_1^{\alpha} + S_2^{\alpha})^2$ for the $S=\frac{1}{2}$ ladder
corresponds to the operator $(S^{\alpha})^2$ in the $S=1$ chain and we
therefore identify sector B with the $XY2$ phase for the $S=1$ chain
as described in ref.[\onlinecite{Schulz1}].

With increasing interleg ferromagnetic coupling
we reach the line $\Delta = |J_{\bot }^{z}|/2J$ which marks
the transition into the phase where both fields are gapless.  In the
sector C of the phase diagram the system shows properties of two {\em
almost independent} spin $S=1/2$ anisotropic Heisenberg chains with
dominating ferromagnetic coupling. The transverse correlations in this
phase are given by
\bea\label{CorrelxySPLQII}
K^{xy}_{\alpha\beta}(r) & \simeq & \delta_{\alpha\beta}
\Big[r^{-1/4(1/K^{\ast}_{+} + 1/K^{\ast}_{-})}\nonumber\\ 
& + & (-1)^{r}\cdot r^{-\left( K^{\ast}_{+} + K^{\ast}_{-} + 1/4K^{\ast}_{+} + 1/4K^{\ast}_{-}\, 
\right) }\Big],
\eea
where $\delta_{\alpha, \beta}$ is the Kronecker symbol.
The longitudinal correlations decay faster. In particular the intraleg
longitudinal correlations are given by
\be\label{CorrelzzSPLQIIa}
K^{zz}_{\alpha\alpha}(r) \simeq \frac{K^{\ast}_{+} + K^{\ast}_{-}}{2\pi r^{2}}
+(-1)^{r}\cdot\, r^{-\left(K^{\ast}_{+} + K^{\ast}_{-}\right)}\, .
\ee
The transverse interleg correlations are strongly suppressed in this
phase, while the longitudinal part of the interleg spin-spin
correlations is given by
\be\label{CorrelzzSPLQIIb}
K^{z}_{\alpha\beta}(r) \simeq \frac{K^{\ast}_{+} - K^{\ast}_{-}}{2\pi r^{2}}
\ee
This phase we denote as the {\em spin liquid II} phase. 

Although the analysis as considered above is formally valid in the
weak-coupling limit ($\Delta, |J_{\bot}^{z}| \ll J$) we can estimate
the upper boundary for the {\em spin liquid II} phase, in the vicinity
of the single chain ferromagnetic instability regime using the
dimensionality analysis. We determine the instability curve
corresponding to the transition into the gapped phase from the
condition $K_{\pm}=1$, where the scaling dimension of the
corresponding {\em cosine} term $d_{\pm} = 2K_{\pm} = 2$. After some
simple algebra one easily obtains that at $J^{z}_{\bot} > 0$ the field
$\phi_{-}$ is gapless, while the field $\phi_{+}$ becomes massive for
\be\label{Jc+Critical}
J_{\bot}^z > J_{+}^{c} = 2 \pi u \frac{K-1}{K^{2}}\, . 
\ee
For $\Delta = 1 -\epsilon$ with $\epsilon \ll 1$, which implies 
$1/K  \sim \sqrt{2\epsilon} \ll 1$ equation (\ref{Jc+Critical}) takes
the following form 
\be\label{JcFinal}
J_{+}^{c}(\Delta) = 
         4 J \epsilon \left(1-\frac{2\sqrt{2\epsilon}}{\pi}\right)\, .
\ee
Therefore in the vicinity of the single chain ferromagnetic
instability point, at $1-\Delta \ll 1$, the {\em spin liquid I} phase
with only one gapless (here antisymmetric) mode reenters the phase
diagram at $J_{\bot}^z > J_{+}^{c}(\Delta)$.  (We note that the
amplitude of the $cosine$ term in the limit of the single chain
ferromagnetic instability point is not determined exactly, so the
phase transition line determined by the dimensional analysis is of
qualitative nature in this limit). For $J^{z}_{\bot} < 0$ the
analysis is done in exactly the same manner with symmetric and
antisymmetric modes changing roles. 

At $J_{\bot}^{z}=0$ and $\Delta >1$ each of the decoupled legs is
unstable towards the transition into a ferromagnetic phase. At
$J_{\bot}^{z} \neq 0$, we can address the problem of the ferromagnetic
instability in the ladder system studying the velocity renormalization
of the corresponding gapless excitations. In analogy with the single
chain case we mark the transition into the ferromagnetically ordered
phase at $u_{\pm}=0$. Using Eqs. (\ref{u}) one finds that the
ferromagnetic transition takes place at
\be\label{JcFinal1}
J^{z}_{FM}(\Delta) = 4J\epsilon\, .
\ee

At $|J_{\bot}^{z}| \gg J$ the boundary of the ferromagnetic
instability can be established from the large rung coupling expansion
approach. Let us first consider the case of strong ferromagnetic
intrarung interaction $J_{\bot}^{z} < 0$. In this limit a large gap of
order $|J_{\bot}^{z}|$ exists in the one-magnon excitation
spectrum. Projecting the system on the subspace excluding antiparallel
orientation of spins within a given rung, in the second-order
perturbation expansion withr respect to $J^{2}/|J_{\bot}^{z}|$ and up
to the additive constant $E_{0}=-N_{0}|J_{\bot}^{z}|$ we obtain the
following effective spin-1/2 $XXZ$ spin chain Hamiltonian
\be
{\cal H}= \sum_{n}\left[\frac{1}{2}\lambda_{eff}^{xy}
(\tau^{+}_{n}\tau^{-}_{n+1}+h.c.) + \lambda^{z}_{eff}\tau^{z}_{n}\tau^{z}_{n+1}\right],
\label{spinhamilt}
\ee
where 
\be
\lambda^{xy}_{eff}  = - \frac{J^{2}}{|J_{\bot}^{z}|}, \quad 
\lambda^{z}_{eff}  =  \frac{J^{2}}{|J_{\bot}^{z}|} - 2J\Delta
\ee
and the pseudospin operators are  
\begin{eqnarray}
\tau^{+}_{n}& = &  S^{+}_{n,1}S^{+}_{n,2}\, , \quad \tau^{-}_{n} =  S^{-}_{n,1}S^{-}_{n,2}\, ,
\nonumber\\  
\tau^{z}_{n}& = &{1\over 2}( S^{z}_{n,1}+S^{z}_{n,2}) \nonumber \, .
\end{eqnarray}

In agreement with the weak-coupling bosonization analysis, at
$\Delta=0$ ($XY$-legs) the system is equivalent to the $S=1/2$
isotropic antiferromagnetic chain. For arbitrary $\Delta < 0 $
($\lambda^{z}_{eff} > \lambda^{xy}_{eff})$ the spin chain given by
the Hamiltonian (\ref{spinhamilt}) is in the {\em gapped N\'eel
phase}. This phase corresponds to the LRO AFM interleg ordering with
interleg phase shift equal to zero. At 
$$ 0 < \Delta <J/|J_{\bot}^{z}| $$ 
$(-\lambda^{xy}_{eff} < \lambda^{z}_{eff} <\lambda^{xy}_{eff})$ the
spin chain (\ref{spinhamilt}) is in a {\em gapless planar} $XY$ phase,
corresponding to the ''spin liquid I'' phase of the bosonization
studies and finally at
$$ \Delta > J/|J_{\bot}^{z}| $$ 
$(\lambda^{z}_{eff}< -\lambda^{xy}_{eff})$ the transition into the
completely polarized ferromagnetic phase takes place.

In the case of strong antiferromagnetic interleg coupling
$J_{\bot}^{z} \gg J >0$ analysis is similar. In this case the intrarung
ordering of spins is antiferromagnetic. Projecting the system on the
subspace excluding {\em parallel} orientation of spins within the same
rung, and introducing a new set of spin operators
\begin{eqnarray}
\tilde{\tau}^{+}_{n}& = &  S^{+}_{n,1}S^{-}_{n,2}\, , \quad \tilde{\tau}^{-}_{n}=S^{-}_{n,1}S^{+}_{n,2}
 \, , \nonumber\\  
\tilde{\tau}^{z}_{n}& = & {1\over 2}(S^{z}_{n,1}-S^{z}_{n,2})\, , \nonumber
\end{eqnarray}
in the second-order with respect to $J^{2}/J_{\bot}^{z}$ we once again
map the initial ladder model onto the theory of an anisotropic spin
$1/2$ Heisenberg chain (\ref{spinhamilt}). One can perform the
analysis as discussed above, however the ferromagnetic ordering in
terms of the effective $S=1/2$ chain, at $J_{\bot}^{z}>0$ corresponds
to an interleg ferromagnetic ordering with a phase shift of $\pi$ of
the order parameter along the rung.

\begin{figure}
\centerline{\psfig{file=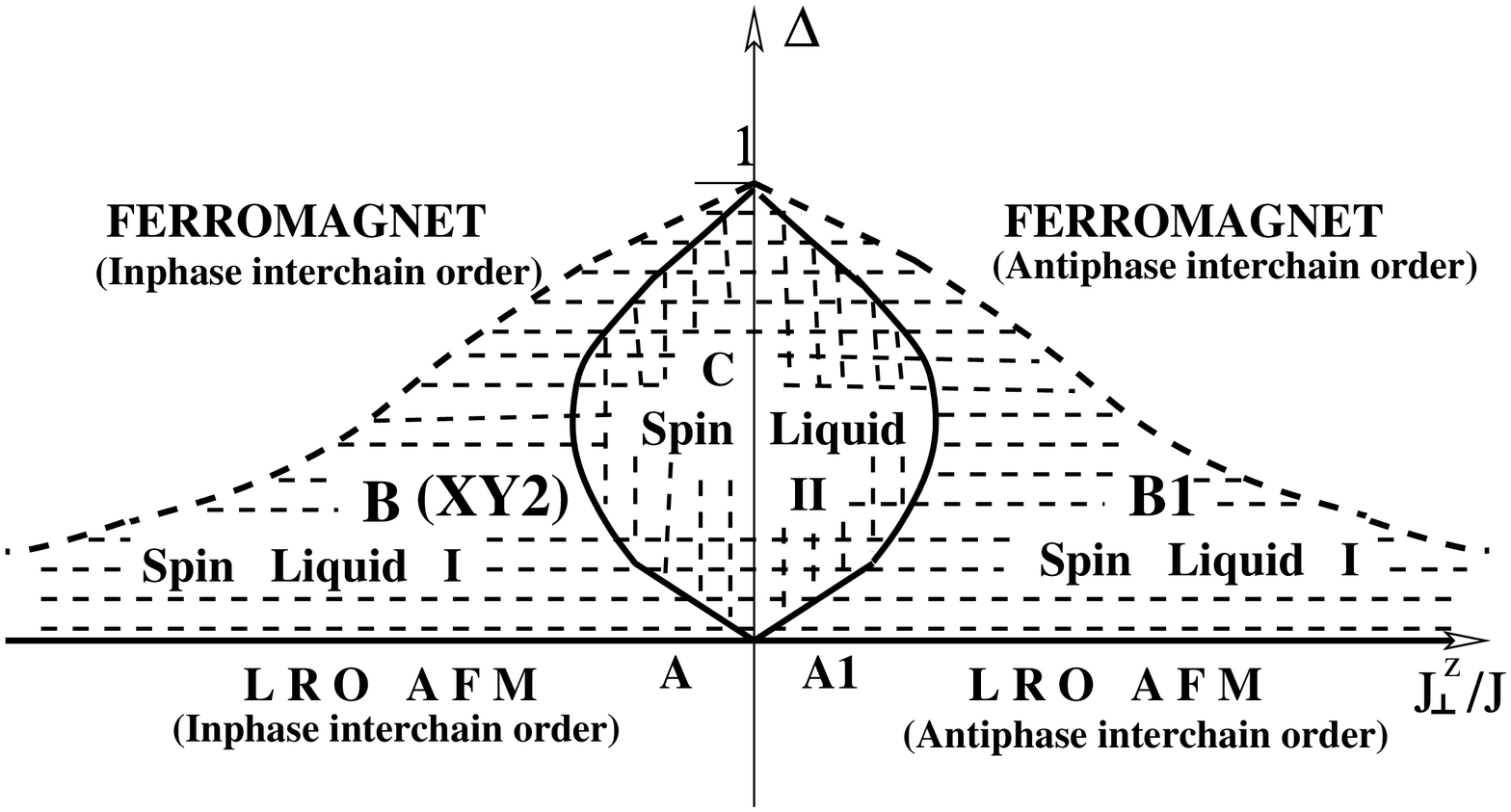,width=85mm,silent=}}
\vspace{10mm}
\caption{The groundstate phase diagram of the two-leg ladder with a longitudinal 
$J^{z}_{\bot}S^{z}_{j,1}S^{z}_{j,2}$ coupling between legs. For details
see text.}
\label{fig:PDzzV1}
\end{figure}

The results obtained within the bosonization approach together with
the results from the strong coupling expansion allow to draw the
following phase diagram of the ladder with a longitudinal interleg
coupling $J^{z}_{\bot}$ (see Fig.\ref{fig:PDzzV1}). At $\Delta < 0$
the phase diagram consists of two gapped phases describing
respectively long range ordered N\'eel antiferromagnetic phases with
gapped excitation spectrum and inphase (at $J^{z}_{\bot}<0$) or
(antiphase at $J^{z}_{\bot}>0$) ordering of spins within the same
rung.  The line $\Delta=0$ marks the transition into the {\em Spin
Liquid I} - phase characterized by a gapless excitation spectrum and
power law decay of the spin-spin correlation functions. The critical
indices for the decay of the corresponding spin-spin correlations in
the Spin Liquid I phase are $\gamma_{i} \simeq 1$. In the case of
strong interleg exchange $|J^{z}_{\bot}| \gg J$, further increase of
the interleg ferromagnetic exchange $\Delta$ leads to the transition
at $\Delta_{c} \simeq J/|J^{z}_{\bot}|$ into the phase with
ferromagnetically ordered legs. However, in the weak-coupling case, at
$|J^{z}_{\bot}| \ll J$, an increase of the parameter $\Delta$ at given
$J^{z}_{\bot}$ leads to the transition into the {\em Spin Liquid II}
at $\Delta_{c(1)}=|J^{z}_{\bot}|/2J$. The Spin Liquid II phase is
characterized by a gapless excitation spectrum and power-law decay of
the spin-spin correlation functions with critical indices $\gamma_{i}
\simeq 2$. This transition marks the development of the regime
dominated by intraleg coupling, whereas the interleg longitudinal
exchange plays only a rather moderate role. However, with further
increase of the intraleg ferromagnetic exchange, in the vicinity of
the ferromagnetic instability line the {\em Spin Liquid II} phase
becomes unstable and the system reentres into the {\em Spin Liquid I}
phase. This reentrance effect is connected with a sharp reduction of the
bandwidth in the vicinity of the ferromagnetic transition and
a subsequent increase of the potential energy of the interleg
coupling. Therefore, just before the transition into the
ferromagnetically ordered phase, the short range interleg fluctuations
get stopped, and as in the case of the strong inrtarung coupling, the
{\em Spin Liquid I} - phase is unstable toward the transition into the
phase with ferromagnetically ordered legs.

However, since the transition into the ferromagnetic phase is a
typical finite bandwidth effect, the parameters determined
quantitatively within the bosonization (i.e. infinite band) approach
strongly depend on the way of regularization of the continuum theory
on small distances. Therefore, it is useful to determine the lowest
boundary of the ferromagnetic phase on the phase diagram, starting
from the ferromagnetically ordered phase and using the standard
spin-wave analysis (see appendix).  At $|J_{\bot}^{z}| \ll J$ this
approach gives $J^{SW}_{FM} = 2 J \epsilon$. This discrepancy clearly
is the result of the linearized expressions used for the parameters
$K_{\pm}$, Eq. (\ref{Kpm}). However, as long as the multiplicative
renormalization used here for the parameters $u_{\pm}$ and $K_{\pm}$,
Eqs. (\ref{upm})-(\ref{Kpm}) remains valid, the scenario discussed
above, where the {\em spin liquid I} phase is unstable towards the
transition into the phase with ferromagnetically ordered legs, remains
qualitatively plausible. For a more quantitative description,
sufficiently detailed numerical studies of this sector of the phase
will be very helpful.

To conclude this subsection we note that the ground state phase
diagram of the ferromagnetic ladder system coupled only by the
longitudinal part of the spin-spin exchange interaction exhibits a
rather rich phase diagram which consist of LRO AFM phases, a spin
liquid phase with one gapped and one gapless mode, a spin liquid phase
with two gapless modes and a phase with ferromagnetically ordered
legs.

\subsection{Chains coupled by the transverse part of the ladder exchange}

In this subsection we consider the case of two critical Heisenberg
chains coupled by a transverse interleg exchange interaction
$J^{z}_{\bot}=0$ and $J^{xy}_{\bot} \neq 0$. The particular aspects of
this limiting case are the following ones:
\begin{itemize}
\item The antisymmetric mode is gapped at arbitrary $J^{xy}_{\bot} \neq 0$.
\item The low energy properties of the system are determined by the
behavior of the symmetric field.
\item The infrared properties of the symmetric field are determined by
the subtle coupling between the symmetric and antisymmetric modes.
\end{itemize}
We start our analysis from the limiting case of weakly anisotropic
$XY$ chains, coupled by the weak interleg transverse exchange, assuming
$|\Delta|, |{\cal J}^{xy}_{\bot}|/J \ll 1$. At $J^{xy}_{\bot} \neq 0$
the antisymmetric mode is gapped and the dual antisymmetric field is
''pinned'' with vacuum expectation value

\begin{equation}  \label{ORDERFIELDS3}
\langle \theta_{-}\rangle  = \left\{ 
\begin{array}{l}
\sqrt{\pi K_{-}/2} \hskip0.5cm \textrm{at $J_{\bot}^{xy} > 0$} \\ 
0 \hskip1.8cm \textrm{ at $ J_{\bot}^{xy} < 0$}
\end{array}
\right. \, .
\end{equation}
Behavior of the symmetric field is governed by the SG Hamiltonian
(\ref{SGeffective}). The standard RG analysis gives that the symmetric
mode is gapped at
\be
\Delta < \Delta_{c1} = \frac{\gamma}{4 J}\cdot |{\cal J}_{+-}| \, . 
\ee

Therefore at $\Delta < \Delta_{c1}$ the excitation spectrum of the
system is gapped. The dynamical generation of a gap in the symmetric
mode leads to condensation of the field $\phi_{+}$ with a vacuum
expectation value $\left < \phi_{+}\right >= 0$. Since the dual
component of the antisymmetric field is ''pinned'' with vacuum
expectation value given by ~(\ref{ORDERFIELDS3}), the so-called
''disordered'' phase \cite{Schulz1} is realized in the ground
state. At $J_{\bot}^{xy} > 0$, spins on the same rung form a singlet and
the ground state corresponds to the state with a singlet pair on each
rung. There is no correlation between spins along the ladder.  In the
case of ferromagnetic coupling, at $J_{\bot }^{xy} < 0$ spins on the
same rung form a state corresponding to the $S^{z}=0$ component of the
triplet (an ''asymmetric triplet'' pair) and the ground state
corresponds to the state with an ''asymmetric triplet'' pair on each
rung. In analogy to the phases of the $S=1$ chain as discused in
\cite{Schulz1} we denote this phase as ''anisotropic large D phase''.
\begin{figure} \vspace{10mm}
\centerline{\psfig{file=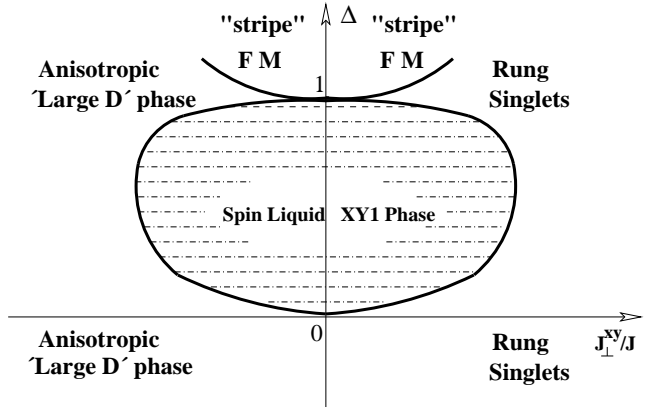,width=85mm,silent=}} \vspace{5mm}
\caption{The groundstate phase diagram of the two-leg ladder with
transverse coupling between legs.}  \label{fig:JxyGSPD} \end{figure}

For $\Delta \geq \Delta_{c1}$ the system is in the phase where the
symmetric mode is gapless. Since the antisymmetric mode is gapped the
{\em alternating part} of the spin-spin correlations is {\em
exponentially small}, while the {\em smooth part shows a power law
decay} at large distances. In particular, the in-plane correlations
are given by
\be\label{CorrelationXY1xySM}
K^{xy}_{\alpha\beta}(r)  \simeq  {\it constant}\cdot 
\left(-\frac{J^{xy}_{\bot}}{|J^{xy}_{\bot}|}\right)^{\alpha + \beta} r^{-1/4K^{\ast}_{+}}\, 
\ee
while the longitudinal correlations decay faster
\be\label{CorrelationXY1zzSM}
K^{zz}_{\alpha\beta}(r) \simeq \frac{ K^{\ast}_{+}}{2\pi r^{2}}\, .
\ee

As follows from Eq. (\ref{CorrelationXY1xySM}) the line
$J_{\bot}^{xy}=0$ marks the transition from a regime with
ferromagnetic interleg order into the regime with antiferromagnetic
interleg order. In the vicinity of this critical line
$J_{\bot}^{xy}=0$ the gap in the antisymmetric mode ($M_{-}$) is tiny,
therefore the correlation functions given by
Eqs. (\ref{CorrelationXY1xySM})-(\ref{CorrelationXY1zzSM}) are valid
only for distances $r \gg L_{c} \sim 1/M_{-}$. However, at distances
$r \ll L_{c}$ fluctuations of the antisymmetric mode are strong, and
the behavior of correlation functions is the same as in the {\em spin
liquid II} phase Eqs. (\ref{CorrelxySPLQII})-(\ref{CorrelzzSPLQIIb}).
Following Schulz \cite{Schulz1} who has discussed a similar phase in
the context of the spin $S=1$ chain we denote this phase as a {\em
spin liquid XY1} phase.

Let us now discuss the phase diagram of the model in the vicinity of
the single chain ferromagnetic instability point $\Delta=1$. As
$\Delta \rightarrow 1$, the effective coupling constant behaves as
$$ \frac{M^{+}_{eff}}{2\pi u} \simeq 
     \frac{\pi^{2} K {\cal J}_{+-}}{2J} \sim \frac{1}{\epsilon} $$
where $\epsilon = 1- \Delta$. A rough estimate of the renormalization
of the velocity of the symmetric mode excitations $u_{+}$ and of the
spin-liquid parameter $K_{+}$ at $\Delta \simeq 1$, in second order
gives
\bea
u_{+} &= & u\left(1 + \lambda K \left({\cal J}_{+-}/J \right)^{2}\right)\, ,
\label{u+tr}\\
K_{+} & =&  K\left(1 - \lambda K \left({\cal J}_{+-}/J \right)^{2}\right)\, ,
\label{K+tr}
\eea
where $\lambda $ is a nonuniversal constant of the order of unity. As
follows from Eqs. (\ref{u+tr}/\ref{K+tr}) the strong effective
transverse coupling reduces the tendency towards ferromagnetic
ordering and leads to the transition into the gapped phase at
$$  {\cal J}_{+-} > 
    {\cal J}_{+-}^{cr} \simeq  J(1-\triangle)^ {1\over 4}.$$  
Equivalently we have
$$\Delta_{c2} \simeq 1-\lambda^2({\cal J}_{+-}/J)^4.$$
To summarize this subsection we note that the ground state phase
diagram of the ferromagnetic ladder system coupled by the transverse
part of the spin-spin exchange interaction only also exhibits a rich
phase diagram which consist of the ''disordered rung-singlet'' and
''anisotropic large D'' phases, the easy-plane gapless XY1 phase and
the ''stripe'' ferromagnetic phases with dominating intraleg
ferromagnetic ordering.

\subsection{Chains coupled  by isotropic interleg exchange}

In this subsection we consider the weak-coupling ground state phase
diagram of the model ~(\ref{Hamiltonian}) in the case of an isotropic
interleg exchange $J^{z}_{\bot}=J^{xy}_{\bot} =: {\cal J}_{\bot}$.

In this case the behavior of the antisymmetric sector is completely
similar to the above considered case of the ladder with transverse
exchange: the antisymmetric field is gapped and the vacuum expectation
value of the dual field $\theta_{-}$, depends on the sign of exchange
and is given by Eq. (\ref{ORDERFIELDS3}) after the substitution ${\cal
J}^{xy}_{\bot} \rightarrow {\cal J}_{\bot}$.

The far-infrared properties of the symmetric field are governed by the effective SG Hamiltonian 
(\ref{SGeffective}) with the bare values of the model parameters given by $K_{+}$ and 
\bea
K_{+} & = & K \, \left(1-\frac{{\cal J}_{\bot} K}{2 \pi u} \right)\, ,
\nonumber\\
M^{+}_{eff} & = & - \frac{1}{\pi u_{+}}{\cal J}_{\bot} \left(1+ \delta \right) \, ,
\eea
(where $\delta$ is a nonuniversal positive number)
\par
The resulting asymmetry of the model is clearly seen: 
\begin{itemize}
\item{at ${\cal J}_{\bot} > 0$, the antiferromagnetic interleg
exchange reduces $K_{+}$ and increases $M^{+}_{eff}$ and therefore
supports the tendencies towards development of a gap in the excitation
spectrum}
\item{at ${\cal J}_{\bot} < 0$, the ferromagnetic interleg exchange
increases $K_{+}$, while with increasing $|{\cal J}_{\bot}|$ the
parameter $M^{+}_{eff} \simeq {\cal J}_{\bot}\left(1 - \delta \right)
\rightarrow 0$ ; therefore we expect an enlargement of the gapless
section in this case.}
\end{itemize}

We start our analysis from the limiting case of weakly anisotropic
$XY$ chains assuming $|\Delta|,|{\cal J}_{\bot}|/J \ll 1$. At $\Delta
= 0$ we have $K=1$ and the system shows a gap in the excitation
spectrum at ${\cal J}_{\bot} > 0$ and is gapless in the case of
ferromagnetic interleg exchange ${\cal J}_{\bot} < 0$. Therefore at
$\Delta = 0$, with increasing ferromagnetic interleg exchange (${\cal
J}_{\bot} < 0,\, |{\cal J}_{\bot}| \rightarrow \infty$) the system
continuously evolves into the limit of the $S=1$ $XY$ model, which is
known to be gapless \cite{Kubo,Nomura}. In the case of
antiferromagnetic interleg exchange ${\cal J}_{\bot} > 0$ the
symmetric mode is unstable towards the Kosterlitz-Thouless type
transition associated with the dynamical generation of a gap in the
excitation spectrum. The weak-coupling RG analysis tells, that at
$\Delta \neq 0$ and ${\cal J}_{\bot} > 0$ the gapless $XY1$ phase is
realized for
\begin{equation}\label{JcRG}
\Delta > \Delta_{c1} = \frac{{\cal J}_{\bot}}{2J}\left(1 + \delta \right)\, ,  
\end{equation} 
whereas in the case of ferromagnetic interrung exchange ${\cal J}_{\bot}
< 0$ it is realized for
$$ \Delta > \Delta^{\prime}_{c1} = -{\delta {\cal J}_{\bot}\over 2J
}\, .  $$

Therefore, from the RG studies we obtain that the gapless $XY1$ phase
is stable in the case of ferromagnetic exchange. At ${\cal J}_{\bot} >
0$ it is unstable towards the transition into the gapped rung-singlet
phase. At ${\cal J}_{\bot} < 0$ the gapless $XY1$ phase penetrates
into the $\Delta <0$ sector of the phase diagram. However, since
$M^{+}_{eff} \rightarrow 0$ with increasing ferromagnetic exchange, at
$|{\cal J}_{\bot}| \gg J$ the gapless phase on the antiferromagnetic
side ($\Delta <0$) of the phase diagram shrinks up to a narrow stripe
which exponentially disappears as $|{\cal J}_{\bot}|/J \to \infty$.

With $\Delta \to 1$ the gapless $XY1$ phase becomes unstable towards
transition into the ferromagnetically ordered state. Following the
route developed before, we find that at $\Delta = 1- \epsilon $ and
antiferromagnetic interleg exchange, ${\cal J}_{\bot} > 0$, the
reentrance of the gapped rung-singlet phase takes place at
\be\label{JcFinal2}
\Delta_{c2} = 1 - \frac{{\cal J}_{\bot}}{4J} + 
              {\cal O}\left(\frac{{\cal J}_{\bot}}{4J}\right)^{3/2}\, .
\ee

Thus, in agreement with the quasiclassical studies
\cite{KolezhukMikeska}, we obtain that two almost ferromagnetically
ordered chains coupled by an isotropic interleg exchange are unstable
towards formation of the gapped rung-singlet phase at $J_{\bot} >
J^{c}_{\bot} > 0$, where $J^{c}_{\bot} \rightarrow 0$ as $\Delta
\rightarrow 1$. However, in contrast to the quasiclassical case,
${\cal J}^{c2}$ increases linearly with $\epsilon$ in the quantum
spin-ladder case.

In the case of ferromagnetic interleg exchange, ${\cal J}_{\bot} < 0$,
the gapless $XY1$ phase becomes unstable towards the transition into
the ferromagnetically ordered phase when $\Delta$ increases towards
1. In this case the spin-wave approach (see appendix) gives that the
boundary between the $XY1$ and the ferromagnetic phase is $\Delta = 1$.

We summarize our results considering the phase diagram of the ladder
with ferromagnetic legs and an isotropic interleg exchange in
Fig. \ref{Fig:JzGSPD}.
\begin{figure}
\vspace{10mm}
\centerline{\psfig{file=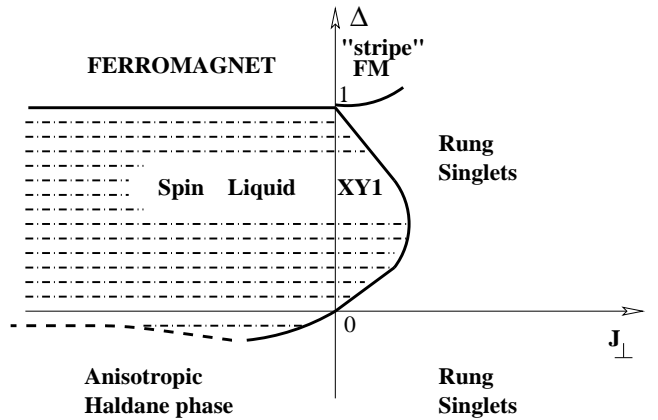,width=85mm,silent=}}
\vspace{5mm}
\caption{The ground state phase diagram of the two-leg ladder with an isotropic interleg coupling.}
\label{Fig:JzGSPD}
\end{figure}

\section {Conclusions}

We have studied the ground state phase diagram of the $S=1/2$ ladder
with ferromagnetically interacting legs using the continuum limit
bosonization approach. The phase diagrams for the extreme anisotropic
interchain coupling cases (Ising and $XY$ interleg exchange) as well
as for the $SU(2)$ symmetric case were obtained. These phase diagrams
exhibit a number of interesting phases, gapped as well as gapless;
some of these are familiar from well known 1D models (rung singlet
phase, anisotropic Haldane phase, ferromagnetic and large $D$ phase),
in addition we describe here for the first time less conventional
phases for ladders: the spin liquid phases with (i) one gapless and
one gapped mode (including the known $XY1$ and $XY2$ phases) and (ii)
two gapless modes. We have shown moreover that the gapped rung singlet
phase found semiclassically to appear for an arbitrarily small
isotropic antiferromagnetic interaction between ferromagnetic legs
\cite{KolezhukMikeska} continues to exist for $S=1/2$ ladders and
$xy-$like interactions and actually extends to small values of
$\Delta$.

The neighborhood of the single chain ferromagnetic instability point
turned out to be of particular interest. We investigated the behavior
of the system in this regime using the multiplicative regularization
scheme. This scheme allows to extend the bosonization formalism to
the limit when the bandwidth of the single chain excitations collapses
and leads to the result that upon increasing the strength of
ferromagnetism $\Delta$ at any moderate fixed interleg interaction a
sequence of two phase transitions occurs before the system enters the
final ferromagnetically ordered phase.

Preliminary investigations show that the system considered here
displays additional interesting aspects when an external magnetic
field (both longitudinal and transverse) is applied. These
investigations will be reported in a subsequent publication.


\section{Acknowledgments}
It is a pleasure to thank A.K. Kolezhuk for helpful discussions. TV is
grateful to A.A. Nersesyan for various enlightening discussions. GIJ
thanks W. Brenig and A. Honecker for interesting discussion. He also
acknowledges support by the SCOPES grant N 7GEPJ62379. This work is
supported by the DFG-Graduiertenkolleg No. 282, ''Quantum Field Theory
Methods in Particle Physics, Gravitational Physics, and Statistical
Physics''.


\appendix 

\section{Ferromagnetic instability}

In this appendix we consider {\em ferromagnetic} interleg coupling,
assuming $J^{xy}_{\bot},J^{z}_{\bot} < 0$, and use the spin wave
approach to determine the critical line corresponding to the {\em
ferromagnetic instability} in our system. For this purpose we start
from the region of the phase diagram where we can safely assume that
the ground state is the fully polarized ferromagnetic state (that is
$\Delta \gg 1$ and $J_{\bot} < 0$). We identify the transition line
from the fully polarized ground state to some other state as the line
of instability in the spin wave excitation spectrum.

Let us denote the eigenstate of the Hamiltonian (\ref{Hamiltonian})
corresponding to the fully polarized (along the $Z$ axis)
ferromagnetic state by $\left| 0 \right >$. Then
$$ S_{j,\alpha}^{+} \left| 0 \right > = 0, 
              \qquad \mbox{for arbitrary j and} \quad \alpha . $$
It is straightforward to obtain that $\hat{H}\left|0\right > =
E_{0}\left|0\right>$ where
$$ E_{0}= - {N \over 4}\left(\left|J^{z}_{\bot}\right| 
                    +  2J \Delta \right). $$

To construct the lowest excitations in the ferromagnetic phase we
act on the ground state configuration by the spin lowering operator
$S_{n,\alpha}^{-}$. Let us denote by $\left| 1 \right >_{n}$ ($\left|
2 \right >_{n}$) the state obtained by action of the spin lowering
operator on the $n^{\rm th}$ site of leg 1 (leg 2):

$$
\left| 1 \right >_n \equiv S_{n,1}^{-}\left| 0 \right >,  \quad 
\left| 2 \right >_n \equiv S_{n,2}^{-}\left| 0 \right >\, .
$$  

It is straightforward to solve the coupled system of equations of
motion in the subspace $S^z_{\rm tot} = N -1$ and to obtain the following
two sets of excitation frequencies:

\begin{eqnarray}
\omega^{-}(q) & = & -J \cos q + J \Delta - {1 \over 2}\left(J^{z}_{\bot} - J^{xy}_{\bot}  \right )\\
\omega^{+}(q) & = & -J \cos q + J \Delta - {1 \over 2}\left(J^{z}_{\bot} + J^{xy}_{\bot}\right )\, .
\end{eqnarray}

For {\em ferromagnetic interleg exchange} ($J^{z}_{\bot},J^{xy}_{\bot}<0$) we
have
$$
\omega^{-}(q) < \omega^{+}(q)
$$
and from the instability condition $\omega^{-}(q=0)=0$ we obtain
\be
\Delta = 1 + \frac{1}{2J}\left(J^{z}_{\bot}- J^{xy}_{\bot}  \right ) \, .
\ee

In the particular limit of noninteracting chains $(J^{z}_{\bot}=J^{xy}_{\bot}=0)$ as well as 
in the limiting case of the {\em rotationally invariant} interleg coupling 
$(J^{z}_{\bot}=J^{xy}_{\bot})$ the critical line corresponding to the instability 
of the ferromagnetic phase is given by the condition
\be
\Delta = 1  \, .
\ee


\end{document}